\pdfoutput=1
\RequirePackage{ifpdf}
\ifpdf % We~are running pdfTeX in pdf mode
\documentclass[pdftex]{sigma}
\else
\documentclass{sigma}
\fi

\numberwithin{equation}{section}

\begin{document}

\allowdisplaybreaks

\newcommand{\arXivNumber}{2401.00586}

\renewcommand{\PaperNumber}{096}

\FirstPageHeading

\ShortArticleName{Scale Invariant Scattering and Bernoulli Numbers}

\ArticleName{Scale Invariant Scattering and Bernoulli Numbers}

\Author{Thomas L.~CURTRIGHT}

\AuthorNameForHeading{T.L.~Curtright}

\Address{Department of Physics, University of Miami, Coral Gables, FL 33124, USA}
\Email{\href{mailto:curtright@miami.edu}{curtright@miami.edu}}
\URLaddress{\url{https://people.miami.edu/profile/7884b39ba94cfdbc1698d114121a27f5}}

\ArticleDates{Received June 07, 2024, in final form October 14, 2024; Published online October 24, 2024}

\Abstract{Non-relativistic quantum mechanical scattering from an inverse square potential in two spatial dimensions leads to a novel representation of the Bernoulli numbers.}


\Classification{81Q60; 11B68; 11M26}

\section{Introduction}

In this note, I summarize some observations about Bernoulli numbers as
obtained in the context of computing the scattering amplitude for a scale
invariant potential.

In two spatial dimensions non-relativistic scattering by a repulsive
potential $V=\kappa/r^{2}$, with~${\kappa>0}$, results in a surprisingly
simple form for the integrated cross section, $\sigma=\int_{0}^{2\pi}\left(
\frac{{\rm d}\sigma }{{\rm d}\theta}\right) {\rm d}\theta$, when computed using quantum
mechanics. The result for a mono-energetic beam of mass $m$ particles is
\cite{TLC,TLCCV} $\sigma=\frac{2\pi^{2}m\kappa}{\hbar^{2}k}$
where the incident energy is $E=\hbar^{2}k^{2}/( 2m) $. This
result follows from a straightforward application of phase-shift analysis
for the potential $V=\kappa/r^{2}$ upon realizing a remarkable identity
involving the sinc function, $\operatorname{sinc}( z) \equiv\sin(
z) /z$, as shown in \cite{TLC}. For real $x$, a succinct form of the identity in question is\footnote{A selection of other identities of this
type, but not exactly \eqref{RID1} so far as I can tell, can be found in~\cite{Illinois}.}
\begin{equation}
1=\frac{\sin( \pi x) }{\pi x}+2\sum_{l=1}^{\infty}\frac{(
-1) ^{l}\sin\bigl( \pi\sqrt{l^{2}+x^{2}}\bigr) }{\pi\sqrt {l^{2}+x^{2}
}} . \label{RID1}
\end{equation}
All higher powers of $x$ cancel when terms on the right-hand side are expanded as series
in $x^{2}$, as a consequence of familiar $\zeta(2n) $ exact
values for integer $n>0$.

\section{Bessel functions and Bernoulli numbers}

Upon expressing the sinc function in terms of \href{https://en.wikipedia.org/wiki/Bessel_function#Spherical_Bessel_functions:_jn,_yn}
{spherical Bessel functions},
\begin{gather*}
j_{n}( z) =\sqrt{\frac{\pi}{2z}}J_{n+1/2}( z) =(
-z) ^{n}\left( \frac{1}{z}\frac{\rm d}{{\rm d}z}\right) ^{n}\frac{\sin z}{z}
\end{gather*}
one may expand the summand of \eqref{RID1} in powers of $x^{2}$ to obtain
\begin{gather*}
\frac{\sin\bigl( \pi\sqrt{l^{2}+x^{2}}\bigr) }{\pi\sqrt{l^{2}+x^{2}}}
=\sum_{n=1}^{\infty}\frac{\bigl( -\pi x^{2}\bigr) ^{n}}{n!2^{n}}\frac {1}{
l^{n}}\sqrt{\frac{1}{2l}}J_{n+1/2}( \pi l).
\end{gather*}
Performing the sum over $l$ in \eqref{RID1} before the sum over $n$ then
leads to\footnote{Truncating the right-hand side of \eqref{HoHum} to $\sum_{l=1}^{N}$ produces an $O( 1/N) $ error which is largest at
$n=1$.}
\begin{equation}
\frac{\pi^{n}}{\sqrt{2}}=-\frac{( 2n+1) !}{2^{n}n!}\sum
_{l=1}^{\infty}\frac{( -1) ^{l}}{l^{n+1/2}}~J_{n+1/2}( \pi
l)\qquad \text{for integer}\quad n\geq1.  \label{HoHum}
\end{equation}
Note the prefactor on the right-hand side can be expressed in terms of the double
factorial: $( 2n+1) !!=( 2n+1) !/\big( 2^{n}n!\big)
$.

Implicit in \eqref{HoHum} is an identity involving the Bernoulli numbers, as
follows from using series representations for the Bessel functions \cite[Chapter 10, in particular, equation~(10.1.8)]{AandS} and interchanging summations. For even $n$,
\begin{gather*}
\sum_{l=1}^{\infty}\frac{ ( -1 ) ^{l}}{l^{n+1/2}}J_{n+1/2} (
l\pi )\\
\qquad = ( -1 ) ^{ \lfloor n/2 \rfloor }\sum
_{k=0}^{ \lfloor ( n-1 ) /2 \rfloor } ( -1 ) ^{k}
\frac{\sqrt{2} ( 2 \lfloor n/2 \rfloor +2k+1 ) !}{ (
2k+1 ) !\Gamma ( 2 \lfloor n/2 \rfloor -2k ) }\frac{
\zeta ( 2k+2 \lfloor n/2 \rfloor +2 ) }{2^{2k+1}\pi^{2k+2}},
\end{gather*}
while for odd $n$
\begin{gather*}
\sum_{l=1}^{\infty}\frac{ ( -1 ) ^{l}}{l^{n+1/2}}J_{n+1/2} (
l\pi ) \\
\qquad=- ( -1 ) ^{ \lfloor n/2 \rfloor }\sum
_{k=0}^{ \lfloor ( n-1 ) /2 \rfloor } ( -1 ) ^{k}
\frac{\sqrt{2} ( 2 \lfloor n/2 \rfloor +2k+1 ) !}{ (
2k ) !\Gamma ( 2 \lfloor n/2 \rfloor +2-2k ) }\frac{
\zeta ( 2k+2 \lfloor n/2 \rfloor +2 ) }{2^{2k}\pi^{2k+1}}.
\end{gather*}
But then $B_{2n+1}=0$ for $n=1,2,\dots$ and \smash{$\zeta(2n) =\frac{( 2\pi) ^{2n}}{2(2n) !}\vert B_{2n}\vert$} for $n=1,2,\dots$
with the usual phases given by $B_{2n}=( -1) ^{n+1}\vert
B_{2n}\vert $. Thus from \eqref{HoHum} we are led to another, rather
remarkable identity
\begin{equation}
1=( -1) ^{n+1}( 4n+2) \sum_{k=0}^{n}\frac{(2n) !}{n!k!( n-k) !}\left( \frac{B_{n+k+1}}{n+k+1}\right)\qquad
\text{for integer}\quad n\geq1.  \label{RID2}
\end{equation}
Here the sum involves \href{https://en.wikipedia.org/wiki/Trinomial_expansion}
{trinomial coefficients} as well as \href{https://en.wikipedia.org/wiki/Bernoulli_number}
{divided Bernoulli numbers}, $\beta_{m}\equiv B_{m}/m$. It is not
difficult to check the validity of \eqref{RID2} using various expressions of
the Bernoulli numbers as finite, \emph{alternating} sums, e.g., as \href{https://en.wikipedia.org/wiki/Bernoulli_number#Connections_with_combinatorial_numbers}{sums of Worpitzky numbers weighted by the harmonic sequence}.\footnote{An excellent introduction to the
extensive literature on the Bernoulli numbers can be found in \cite{Luschny, Luschny1}.}

\section{A novel representation of Bernoulli numbers}

If encountered as graffiti on the stones of a bridge, e.g., in Ireland,
either \eqref{RID1} or its companion identity \eqref{RID2} might cause
nothing more than a raised eyebrow in passing. Perhaps justifiably~so.

However, upon inverting the linear relations in \eqref{RID2} to obtain
expressions for each individual Bernoulli number, the results are more
striking: The unsigned Bernoulli numbers $\vert B_{2n}\vert $
for $n\geq2$ are given by interesting sums of $n-1$ monotonically decreasing
positive rational numbers. Unlike many other such representations \cite{MathWorld}, here the terms in the finite sums that represent $B_{2n}$ do
\emph{not} alternate in sign.

For example,
\begin{equation}
\left(
\begin{array}{c}
\left\vert B_{2}\right\vert \\[1mm]
\left\vert B_{4}\right\vert \\[1mm]
\left\vert B_{6}\right\vert \\[1mm]
\left\vert B_{8}\right\vert \\[1mm]
\left\vert B_{10}\right\vert \\[1mm]
\left\vert B_{12}\right\vert \\[1mm]
\left\vert B_{14}\right\vert \\[1mm]
\left\vert B_{16}\right\vert \\[1mm]
\left\vert B_{18}\right\vert \\[1mm]
\left\vert B_{20}\right\vert
\end{array}
\right) =\left(
\begin{array}{c}
\frac{1}{6}\\[2mm]
\frac{1}{30}\\[2mm]
\frac{1}{42}\\[2mm]
\frac{1}{30}\\[2mm]
\frac{5}{66}\\[2mm]
\frac{691}{2730}\\[2mm]
\frac{7}{6}\\[2mm]
\frac{3617}{510}\\[2mm]
\frac{43 867}{798}\\[2mm]
\frac{174 611}{330}
\end{array}
\right) =\left(
\begin{array}{l}
\frac{1}{6}\\[2mm]
\frac{1}{30}\\[2mm]
\frac{1}{60}+\frac{1}{140}\\[2mm]
\frac{1}{45}+\frac{1}{105}+\frac{1}{630}\\[2mm]
\frac{1}{20}+\frac{3}{140}+\frac{1}{252}+\frac{1}{2772}\\[2mm]
\frac{1}{6}+\frac{1}{14}+\frac{17}{1260}+\frac{1}{693}+\frac{1}{12 012}
\\[2mm]
\frac{691}{900}+\frac{691}{2100}+\frac{59}{945}+\frac{41}{5940}+\frac {5}{
10 296}+\frac{1}{51 480}\\[2mm]
\frac{14}{3}+ 2  +\frac{359}{945}+\frac{8}{189}+\frac{4}{1287}+\frac{1}{6435
}+\frac{1}{218 790}\\[2mm]
\frac{3617}{100}+\frac{10 851}{700}+\frac{1237}{420}+\frac{217}{660}+\frac{
293}{12 012}+\frac{1}{780}\\[1mm]
\hphantom{\frac{3617}{100}}{}+\frac{7}{145 860}+\frac{1}{923 780}\\[2mm]
\frac{43 867}{126}+\frac{43 867}{294}+\frac{750 167}{26 460}+\frac {6583
}{2079}+\frac{943}{4004}+\frac{1129}{90 090}\\[1mm]
\hphantom{\frac{43 867}{126}}{}+\frac{217}{437 580}+\frac{2}{
138 567}+\frac{1}{3879 876}
\end{array}
\right).  \label{Examples}
\end{equation}
For these examples, in the finite sequence of terms that sum to give $
\vert B_{2n}\vert $ obviously the second number in the sequence
is just $3/7$ times the first. Less obviously, each term in the sequence
for~$\vert B_{2n}\vert $ is greater than the subtotal of all the
smaller terms in that same sequence.

The general result for $\vert B_{2n}\vert $ is obtained by
writing \eqref{RID2} as an infinite matrix equation, $\boldsymbol{1}=
\boldsymbol{M\cdot B}$, where $\boldsymbol{B}$ is an infinite column of the
even index Bernoulli numbers, $\boldsymbol{1}$ is an infinite column of $1$
s, and $\boldsymbol{M}_{m,n}=2( -1) ^{m+1}\binom{2n-1}{m}\binom{
2m+1}{2n}$. Computing the inverse for the triangular matrix $\boldsymbol{M}
$ then gives $\boldsymbol{B=M}^{-1}\boldsymbol{\cdot1}$. The ordered terms
in the sums of \eqref{Examples} are just the unsigned entries in the
corresponding columns of $\boldsymbol{M}^{-1}$ for the $n$-th row. All
terms in a given row of $\boldsymbol{M}^{-1}$ have the same sign. Some
straightforward algebra then leads to
\[
\vert B_{2n}\vert =\frac{( n!) ^{2}}{(
2n+1) !}+\sum_{k=2}^{n-1}n!~q_{n-k-1}( n) \frac{k!(
k-1) }{( 2k+1) !},
\]
where the $l$-th order polynomials $q_{l}(n)$, with $l\geq0$ and
$n\geq l+3$, may be obtained sequentially from
\[
q_{l}( n) =\frac{( -1) ^{l}( n-l-3) !}{
( 2l+3) !( n-2l-3) !}+\sum_{j=0}^{l-1}\frac{(
-1) ^{l+j+1}( n-l-1+j) !}{( 2l+1-2j) !(
n-2l-1+2j) !}~q_{j}( n+j-l).
\]
For example,
\[
q_{0}( n) =\frac{1}{6},\qquad q_{1}( n) =\frac {7}{360
}n-\frac{1}{45},\qquad q_{2}( n) =\frac{31}{15120}n^{2}-
\frac{89}{15 120}~n+\frac{1}{315},\qquad \text{etc}.
\]

\section{Conclusions}

Many things can be said about the entries in $\boldsymbol{M}^{-1}$, such as
those along the diagonal, i.e.,
\[
\bigl\vert \boldsymbol{M}_{n,n}^{-1}
\bigr\vert =\frac{( n!) ^{2}}{( 2n+1) !},
\]
 the first
sub-diagonal, i.e.,
\[
\bigr\vert \boldsymbol{M}_{n\geq2,n-1}^{-1}\bigr\vert =
\frac{1}{6}( n-2) \frac{n!( n-1) !}{( 2n-1)
!},
\]
the second sub-diagonal, i.e.,
\[
\bigl\vert \boldsymbol{M}
_{n\geq3,n-2}^{-1}\bigr\vert =\frac{7}{360}\left( n-\frac {8}{7}\right)
( n-3) \frac{n!( n-2) !}{( 2n-3) !},
\]
 etc. But those things remain to be said later, and not here.\footnote{Some additional details have been worked out
explicitly in T.S.~Van Kortryk, On Bernoulli numbers (private communication).}

It is also possible to relate the terms in the finite monotonic series for $
\vert B_{2n}\vert $ to various partitions of the \emph{infinite}
monotonic series obtain from
\[
\vert B_{2n}\vert =\frac{2(2n) !}{( 2\pi) ^{2n}}\zeta(2n) =\frac{2(2n) !}{( 2\pi) ^{2n}}\sum_{k=1}^{\infty }1/k^{2n}.
\]
 But
that too remains to be discussed later, and not here.

Suffice it to say here that this monotonic finite series representation of $
B_{2n}=( -1) ^{n+1}\vert B_{2n}\vert $ clearly gives a
series of progressively better bounds on $B_{2n}$. Such a series of
constraints on~$B_{2n}$ might be useful to establish bounds on functions
defined as infinite series whose coefficients involve the Bernoulli numbers
\cite{Riesz}. But that remains to be shown.
Finally, as suggested by an anonymous reviewer, the identity \eqref{RID2}
may have an implicit connection to various multi-linear identities for
Bernoulli numbers \cite{Miki2,Miki}, perhaps due in part to the fact that such
``convolution'' identities also appear in
the context of theoretical physics models~\cite{Dunne}.

These pending developments notwithstanding, in closing it seems appropriate
to note the results presented here have already led to the addition of three
entries to the \href{https://oeis.org/}{Online Encyclopedia of Integer
Sequences} (see \href{https://oeis.org/search?q=A368846}{A368846}, \href{https://oeis.org/search?q=A368847}{A368847} and \href{https://oeis.org/search?q=A368848}{A368848}).

\subsection*{Acknowledgements}

I thank P.~Luschny, C.~Vignat and T.S.~Van Kortryk for comments and
discussions. I received financial support from the United States Social
Security Administration.

\pdfbookmark[1]{References}{ref}
\LastPageEnding

\end{document}